\documentclass[prd,aps,preprintnumbers, showpacs, nofootinbib,superscriptaddress,notitlepage]{revtex4-2}
\usepackage{slashed}
\usepackage{array}
\def\l{\left}
\def\r{\right}
\usepackage{multirow}

\usepackage{subfigure}

\usepackage{color}

\usepackage{epsfig}
\usepackage{bbold}

\usepackage{caption}

\usepackage{amssymb,amsthm,amsmath}
\usepackage{textcomp}
\usepackage{color}      
\usepackage{slashed}    
\usepackage{verbatim}
\usepackage[normalem]{ulem} 
\usepackage{rotating}   
\usepackage{multirow}   
\begin{document}
	\title{ 
The SM expected branching ratio for $h \to \gamma \gamma$ and  an excess  for $h \to Z \gamma$
} 
	\author {
		Xiao-Gang He}\email{hexg@sjtu.edu.cn}
	\affiliation{Tsung-Dao Lee Institute,
		Shanghai Jiao Tong University, Shanghai 200240, China} 
	\affiliation{Key Laboratory for Particle Astrophysics and Cosmology (MOE)
\& Shanghai Key Laboratory for Particle Physics and Cosmology,
Shanghai Jiao Tong University, Shanghai 200240, China}
	\author {
	Zhong-Lv Huang}\email{huangzhonglv@sjtu.edu.cn}
\affiliation{Tsung-Dao Lee Institute,
	Shanghai Jiao Tong University, Shanghai 200240, China} 
	\affiliation{Key Laboratory for Particle Astrophysics and Cosmology (MOE)
\& Shanghai Key Laboratory for Particle Physics and Cosmology,
Shanghai Jiao Tong University, Shanghai 200240, China}
	\author {
	Ming-Wei Li}\email{limw2021@sjtu.edu.cn}
\affiliation{Tsung-Dao Lee Institute,
	Shanghai Jiao Tong University, Shanghai 200240, China} 
	\affiliation{Key Laboratory for Particle Astrophysics and Cosmology (MOE)
\& Shanghai Key Laboratory for Particle Physics and Cosmology,
Shanghai Jiao Tong University, Shanghai 200240, China}
	\author {
		Chia-Wei Liu}\email{
		chiaweiliu@sjtu.edu.cn	}
	\affiliation{Tsung-Dao Lee Institute,
		Shanghai Jiao Tong University, Shanghai 200240, China} 
		\affiliation{Key Laboratory for Particle Astrophysics and Cosmology (MOE)
\& Shanghai Key Laboratory for Particle Physics and Cosmology,
Shanghai Jiao Tong University, Shanghai 200240, China}

\date{\today}

	\begin{abstract}
The recent measurements of  $h \to Z \gamma$ from ATLAS and CMS show an excess of the signal strength $\mu_Z = (\sigma\cdot{\cal B})_{\mathrm{obs}}/(\sigma\cdot{\cal B})_{\mathrm{SM}}=2.2\pm 0.7$,  normalized as 1 in   the standard model~(SM). If confirmed, it would be a signal of new physics (NP) beyond the  SM. We study NP explanation for this excess. In general, for a given model, it also affects the process $h \to \gamma \gamma$. Since the measured branching ratio for this process agrees well with the SM prediction, the model is severely constrained. We find that a minimally fermion singlets and doublet extended NP model can explain simultaneously the current data for $h \to Z \gamma$ and $h\to \gamma\gamma$. There are two solutions. Although both solutions enhance the amplitude of $h \to Z \gamma$ to the observed one, in one of the solutions  the amplitude of $h \to \gamma \gamma$ flips sign to give the observ	ed branching ratio. This seems to be a contrived solution although cannot be ruled out simply using branching ratio measurements alone. However, we find another solution that naturally enhances $h \to Z \gamma$ to the measured value, but keeps the amplitude of $h \to \gamma\gamma$ close to its SM prediction. We also comment on the phenomenology associated with these new fermions.

	\end{abstract}

	\maketitle	

\section{Introduction}
\label{Introdction}
\begin{figure}[t]
	\begin{center}
		\includegraphics[width=0.23 \linewidth]{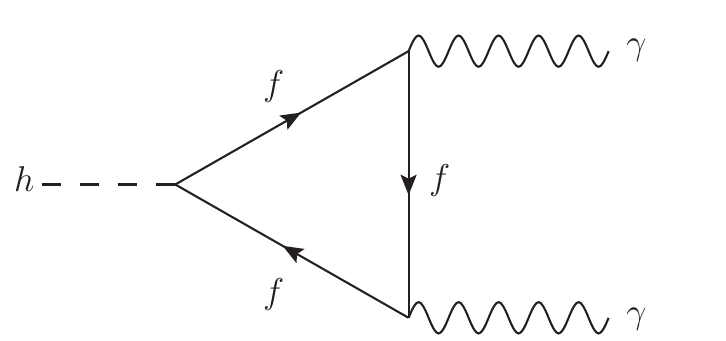}
		\includegraphics[width=0.23 \linewidth]{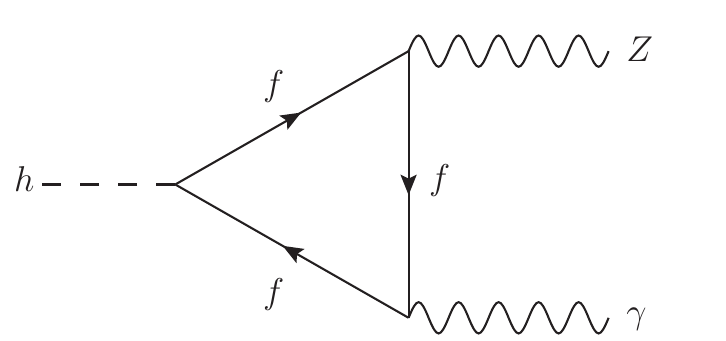}
		\caption{The Feynman diagrams with flavor-conserving vertices.} 
		\label{fig:tri}
	\end{center}
\end{figure}

The 2012 discovery of the Higgs boson~($h$) marked a milestone in particle physics~\cite{ATLAS:2012yve,CMS:2012qbp}. 
Various properties of $h$ predicted by the standard model  (SM) have been confirmed, but there are still many more to be tested.
Notable ones are the $h \to \gamma \gamma$ and  $h \to Z \gamma$~\cite{Ellis:1975ap,Cahn:1978nz,Shifman:1979eb,Gavela:1981ri,Bergstrom:1985hp}. 
They are generated  at loop level in both the SM and new physics~(NP)~\cite{Elias-Miro:2013gya,Dedes:2019bew}, 
making them ideal places to test  NP theories hiding at  loop level.
 In particular,  $h \to \gamma \gamma$ has been measured to good precision, playing a significant role in probing the Higgs boson~\cite{CMS:2013cff,ATLAS:2013dos}.  However, $h \to Z \gamma$ has yet to be confirmed experimentally. 
The   triangle Feynman diagrams induced by fermions are given in FIG.~\ref{fig:tri}. 
 As  they involve the Yukawa couplings of Higgs boson to fermions, the potential influence of NP beyond the SM in these processes is a compelling aspect of ongoing research~\cite{Bizot:2015zaa,Cao:2018cms,Barducci:2023zml,Lichtenstein:2023vza,Hong:2023mwr,Boto:2023bpg,Das:2024tfe,Cheung:2024kml}. 

To gauge how well the SM prediction fits data, it is convenient to define the signal strength, denoted as $\mu = (\sigma\cdot{\cal B})_{\mathrm{obs}}/(\sigma\cdot{\cal B})_{\mathrm{SM}}$. This value represents the observed product of the Higgs boson production cross section ($\sigma$) and its branching ratio (${\cal B}$) normalized by the SM values. When $\mu$ deviates from 1, it is a signal of NP beyond the SM. Recently, both  ATLAS  and CMS~\cite{CMS:2022ahq,ATLAS:2023yqk}  have obtained evidence  of $h\to Z \gamma$ with the average data showing that
\begin{equation}		\label{table:1} 
\mu_Z^{\text{exp}} = 2.2\pm 0.7\,,
\end{equation}
where $\mu_Z$  denotes  the signal strength of $h\to Z \gamma$. The central value  indicates an excess approximately twice as large as that predicted by the SM. On the other hand, the SM prediction aligns very closely with the data of  the
$h\to \gamma \gamma$
signal strength~\cite{ATLAS:2019jst,CMS:2021kom}
\begin{equation}		\label{table:1} 
\mu_\gamma^{\text{exp}} = 1.10\pm 0.07\,.
\end{equation}

At present, the excess for $h \to Z \gamma $ is only at 1.7$\sigma$. If the excess is further confirmed, it will be a signal of NP. In general, for a given NP model addressing the $h \to Z \gamma$ excess, it also affects the process $h \to \gamma \gamma$. Since the measured $\mu _\gamma$ agrees well with the SM prediction, the model is severely constrained. 
For a specific NP model,
the literature notably shows that it is difficult to simultaneously align both $\mu_\gamma$ and $\mu_Z$ with the data within $1\sigma$~\cite{Das:2024tfe,Hong:2023mwr,Lichtenstein:2023vza}.
In this work, we study   the  implications from a  model building point of view for the possible $h \to Z \gamma$ excess problem, the issue  of the SM expected branching ratio for $h \to \gamma \gamma$ and an excess for $h \to Z \gamma$.

Beyond  the SM effects can come in many different ways. 
In the SM effective field theory, the leading    
CP-even 
dimension-six operators     contributing to $h\to \gamma\gamma$ and $h\to  Z \gamma$ are given as~\cite{Korchin:2013ifa}
\begin{eqnarray}\label{eq:006}
	{\cal L}_{\mathrm{eff}} =  c_{BB}\frac{\,{g^\prime}^2}{2 \, \Lambda^2} \,
	H^\dagger \, H \, B_{\mu\, \nu} B^{\mu \, \nu} + 
	c_{WW} \frac{g^2}{2 \, \Lambda^2} \, H^\dagger \, H
	\, W_{a\,\mu\, \nu} W_a^{\mu \, \nu} + 
	c_{WB}\frac{g^\prime \, g}{2 \, \Lambda^2} \,
	H^\dagger \, \tau_a \, H \, W_a^{\mu \, \nu} B_{\mu\, \nu},
\end{eqnarray}
where $H$ is the Higgs doublet with quantum numbers $(1,2, 1/2 )$ under the SM gauge group $SU(3)_c\times SU(2) \times U(1)_Y$, 
and the vacuum expectation value (vev) is given by $\langle H\rangle=(0 ,v/\sqrt{2})^T$ after spontaneous symmetry breaking.
$\Lambda$ represents the energy scale of  NP, $c_{BB,WB,WW}$ are identified as the Wilson coefficients, $\tau_a$ is the Pauli matrix. With the gauge couplings $g'$ and $g$, $B^ {\mu\nu}
= \partial^{\mu} B^ \nu - \partial ^ \nu B ^ \mu
$ and $W^{\mu\nu} _a = 
\partial ^ \mu W_a^\nu    
-\partial ^ \nu W_a^\mu 
+ g \epsilon_{abc} 
W_b^\mu W_c^\nu
$ are gauge field tensors for  $U(1)_Y$ and $SU(2)$ , respectively.
Here we take these operators as examples to show how the excess for  \(\mu_Z^{\text{exp}}\) can be explained. For a full basis that includes CP-violating operators, one may refer to Refs.~\cite{Elias-Miro:2013gya,Pomarol:2013zra}.

After spontaneous symmetry breaking, $B_\mu = A_\mu \cos \theta_W + Z_\mu  \sin \theta_W $ and $ 
W_\mu^3 = Z_\mu \cos \theta_W  - A_\mu \sin \theta_W$, 
we could match the above NP operators  to the effective Lagrangian responsible  to $h\to \gamma \gamma$ and $h\to Z\gamma$ to obtain
\begin{eqnarray}\label{eff_Lag}
	\mathcal{L}_{\mathrm{eff}}^{h \gamma \gamma}=\frac{\alpha_{em} }{8 \pi  v}
( c_{\gamma}^{\text{SM}} + \delta c_{\gamma})
	 F_{\mu \nu} F^{\mu \nu} h\,, \;\;\;\;
	\mathcal{L}_{\mathrm{eff}}^{hZ \gamma}=\frac{\alpha_{em} }{4 \pi  v}
( c_{Z}^{\text{SM}} + \delta c_{Z})
	 Z_{\mu \nu} F^{\mu \nu} h\,,
\end{eqnarray}
with
\begin{eqnarray}\label{eq007}
	\delta c_\gamma &=& 
	\left(\frac{4\,\pi v}{\Lambda}\right)^2\left(c_{BB}+c_{WW} - c_{WB} \right)\,,~~~~\
	\delta c_Z = 
	\left(\frac{4\,\pi
		v}{\Lambda}\right)^2\Bigl(
	\cot\theta_W c_{WW} - 
	\tan\theta_W c_{BB} -\cot2\,\theta_W c_{WB}\Bigr)\,.
\end{eqnarray}
In the above $\theta_W$ is the Weinberg angle, $\alpha_{em}$ is the fine structure constant, $F^{\mu \nu} = \partial^\mu A^\nu - \partial^\nu A^ \mu$ and 
  $Z^{\mu \nu} = \partial^\mu Z^\nu - \partial^\nu Z^ \mu$. 
  
  Including QCD corrections~\cite{Zheng:1990qa,Spira:1991tj,Spira:1995rr, Djouadi:2005gi,Bonciani:2015eua,Gehrmann:2015dua},
  the amplitudes from the SM are effectively encapsulated by 
  $ 
 c_{\gamma}^{\text{SM}}= -6.56$ and $c_{Z}^{\text{SM}} = -11.67$ with $m_h=125.1$~GeV. 
  It is noteworthy that $\delta c_{\gamma}$ and $\delta c_Z$ are influenced by $c_{BB,WB,WW}$ in distinct ways. The experimental values for $c_\gamma$ and $c_Z$ can be accomplished by fine-tuning the NP coefficients to induce a minimal $\delta c_\gamma$ but sizable $\delta c_Z$.
We stress that 
\({\cal L}_{\text{eff}}^{h\gamma \gamma}\) and \({\cal L}_{\text{eff}}^{hZ \gamma}\)  are  $U(1)_{em}$ gauge invariant, and  it suffices to exclusively consider these two
 for our purpose.

From the above analysis, it is clear that by tuning 
$c_{BB}$, $c_{WW}$ and $c_{WB}$ 
 it is possible to simultaneously fit the measured data for $h \to \gamma\gamma$ and the excess in $h\to Z \gamma$. 
 Comprehensive studies on the numerical fit within the framework of the SM effective field theory have been carried out~\cite{Pomarol:2013zra,Dawson:2018pyl,Ellis:2018gqa,deBlas:2022ofj}. 
However, it is still a challenging task to solve the excess problem for $h \to Z \gamma$  with a renormalizable model. 
In a renormalizable model, $\delta c_\gamma$ and $\delta c_Z$ are generated  at one loop level as shown in Figure \ref{fig:tri}.
 In this work, we will focus on constructing a renormalizable model to address the problem.  We find that a minimally extended model with two fermion singlets and a doublet, as shown in Table \ref{rep3}, can explain the $h \to Z \gamma$ problem. 
 We only consider color-singlet fermions because, otherwise, they would lead to dramatic changes in the $gg \to h$ process  and its $p_T$ spectrum, in  contrast to the data~\cite{ATLAS:2022vkf,Battaglia:2021nys,CMS:2024jbe}. 
Two solutions were found. One  suggests  that the SM amplitude $c_Z^{\text{SM}}$ is enhanced by $\delta c_Z$ for $h \to Z \gamma$ to the observed value, however, for $h \to \gamma \gamma$, the decay amplitude $c_\gamma^{\text{SM}} +\delta c_\gamma$ is 
modified 
 to $-c_\gamma^{\text{SM}}$ to give the observed branching ratio. This solution seems to be a contrived solution, although it cannot be ruled out simply using branching ratio measurements. We, however, find another solution which naturally enhances the $h \to Z \gamma$ to the measured value,  while keeping the amplitude of  $h \to \gamma\gamma$ close to its SM prediction.   The model suggests the existence of three new fermions that primarily decay into another fermion and a SM gauge  boson.
Additionally, it proposes a stable fermion with an electric charge close to $-8$ and masses around 2 TeV.

	\begin{table}[b]
		\begin{center}   
			\caption{The fermion representations  in the minimal fermion extension model. The fermions are vector like, having both left- and right-handed components, therefore the model is automatically gauge anomaly free~\cite{Bonnefoy:2020gyh}. }
			\label{rep3} 
			\begin{tabular}{|c|ccc|}   
				\hline
				\hline  &   ~~~$SU(3)_c$ ~~~&  ~~~$SU(2) $~~~ & ~~~$U(1)_Y $~~~ \\   
				\hline
				$(f_S^{Y+1})_{L,R}$ & ${\bf 1}$ & ${\bf 1}$ & $Y +1  $ \\ 	
				$(f_D)_{L,R}$ & ${\bf 1}$ & ${\bf 2}$ & $Y+ 1/2 $ \\		
				$(f_S^{Y})_{L,R}$ & ${\bf 1}$ & ${\bf 1}$ & $Y  $ \\ 	
				\hline   
				\hline
			\end{tabular}   
		\end{center}   
	\end{table}

\section{The model and its interactions}

The new fermions can couple to the Higgs doublet $H$ and can also have bare masses. 
The Yukawa interaction and bare mass terms are given by
\begin{eqnarray}
	\label{Model}
	{\cal L}_{H+M}  &=&  
	-	m_D
	\overline{f}_D f_D 
	- 
	m_S ^{Y} 
	\overline{f}_S ^Y  f_S ^Y 
	-	m_S ^{Y+1} 
	\overline{f}_S ^{Y+1}  f_S ^{Y+1}
	-\left( c_f^Y\overline{f}_D f_S^{Y} H
	+ c_f^{Y+1} \overline{f}_D f_S^{Y+1 } \tilde H
	+ (\text{h.c.})\right) ,
\end{eqnarray}
where $\tilde H = i\tau_2 H^*$.   
Without loss of generality, we choose the chiral basis in this work such that terms of the form $m \overline{f} \gamma_5 f'$ vanish as detailed in Appendix A.
For a general case,  it is also possible to have CP-violating terms of the form  $\overline{f}\gamma_5 f ' H$. In Appendix A, we outline how the analysis can be carried out. In the following, for clarity and simplicity, we use the above Yukawa coupling to demonstrate how the SM expected branching ratio for $h \to \gamma \gamma$ and an excess for $h \to Z \gamma$ can be realized.

The non-zero vev $v$ of Higgs can contribute to fermion masses leading to the new fermion mass matrices in the basis $(f^X_D, f_S^X)^T$  with $X = Y$ or $Y+1$ as the following
\begin{equation}
	\hat M^X = 	 \left(
	\begin{array}{cc}
		m_D & c_f^Xv/\sqrt{2}  \\
		c_f^Xv/\sqrt{2} & m_S^X
	\end{array}
	\right )\,.
\end{equation}
The  eigenstates $f^X = (f_1^X,f_2^X)^T$ of $\hat M^X$ read
\begin{equation}
	\left(
	\begin{array}{c}
		f_1^X \\
		f_2 ^X
	\end{array}
	\right) 
	= \left(
	\begin{array}{cc}
		\cos \theta_X & \sin \theta_X  \\
		-\sin \theta_X & \cos \theta_X
	\end{array}
	\right)
	\left(
	\begin{array}{c}
		f_D^X  \\
		f_S  ^X 
	\end{array}
	\right) \,,
\end{equation}
where the eigenvalues and mixing angles are 
\begin{eqnarray}
	m_{1,2}  ^X &=&  
	\frac{m_D + m^X_S}{2} \pm  \sqrt{ 	\frac{(m_D - m^X_S)^2}{4}
		+ \frac{(c_f^X)^2 v^2}{2}
	}\,,
	~~~
	\sin 2	\theta_X =   \frac{\sqrt{2}  c_f^X v} {m^X_1 -m^X_2   }  
	\,.
\end{eqnarray}
In this  basis, the mass  and  charge-conserved Lagrangian for  $f^X$  are given as
\begin{eqnarray}\label{Lag}
	{\cal L}^X_M &=& -
	\left(
	m_1 ^X
	\overline{f}_1^X f_1^X
	+
	m_2 ^X 
	\overline{f}_2^X f_2^X \right) \,,\nonumber\\
	{\cal L}_h ^X
	&=&			- 
	\frac{ c_f^X h}{\sqrt{2}}
	\overline{f}^X   \left( 
	\sin 2 \theta_X \sigma_z + \cos 2 \theta_X  \sigma_x 
	\right)f^X
	\,,\nonumber\\
	{\cal L}_\gamma^X &=&
	e  X  \overline{f}^X A^\mu \gamma_\mu f^X \,,
	\\
	{\cal L}_{Z} ^X
	& = &e
	\overline{f} ^X
	Z ^\mu \gamma_\mu 
	\left[ 
	- X \tan \theta_W   - \frac{\eta_X}{4 \cos \theta_W \sin \theta_W }
	\left( 1 
	+   \cos 2 \theta_X  \sigma_z 
	-  
	\sin  2  \theta_X \sigma_x 
	\right) 
	\right]
	f^X \nonumber \,,
\end{eqnarray}
where $(\eta_{Y}, \eta_{Y+1}) = (1, -1)$ and $\sigma_{x,y,z}$ are the Pauli matrices operating in the $SU(2)_{L+R}$ rotational space of $f^X$. 
For example, $\overline{f}^X \sigma_x f^X = \overline{f}^X_1 f^X_2 + \overline{f}^X_2 f^X_1$.
On the other hand, the $W$-boson can induce a charge current given by 
\begin{eqnarray}
	{\cal L} _W 
	&=&
	\frac{g}{\sqrt{2}}
	W^+_\mu 
	\Big(
	\cos \theta_{Y+1}  \cos \theta_Y  \overline{f}_1 ^{Y+1} \gamma^\mu  f_1 ^Y
	+
	\sin \theta_{Y+1}  \sin  \theta_Y  \overline{f}_2 ^{Y+1} \gamma^\mu  f_2 ^Y
	\nonumber\\
	&-&  \cos \theta_{Y+1} \sin \theta_Y \overline{f}_1 ^{Y+1} \gamma^\mu  f_2 ^Y
	-
	\cos \theta_{Y} \sin \theta_{Y+1}  \overline{f}_2 ^{Y+1}  \gamma^ \mu f_1 ^Y
	\Big) +(\text{h.c.}) \,.
\end{eqnarray}
As we will see shortly, to explain the data we need a large $Y$, which prevents them from coupling with the SM fermions. Hence, the Lagrangian remains invariant under the $U(1)$ rotation of $f_{1,2}^{Y,Y+1} \to e^{i\theta} f_{1,2}^{Y,Y+1}$, and at least one of the new fermions must be stable. 

\section{Loop-induced $h \to \gamma \gamma$ and $h \to Z \gamma$ in the model}

\begin{figure}[t]
    \begin{center}
        \includegraphics[width=0.23 \linewidth]{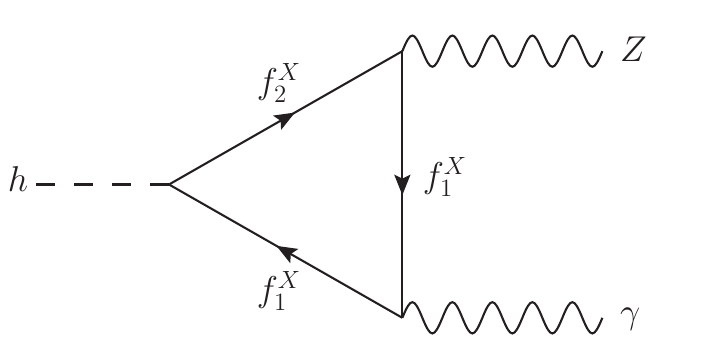}
        \includegraphics[width=0.23 \linewidth]{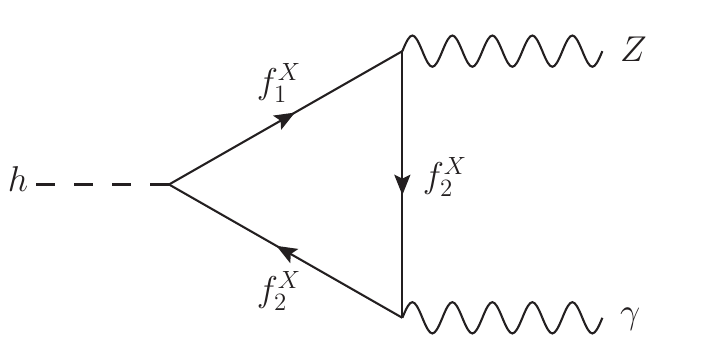}
	\caption{The Feynman diagrams  with flavor-changing vertices}
		\label{fig:triS}
    \end{center}
\end{figure}

We are ready to calculate the loop-induced $h\to \gamma\gamma$ and $h \to Z \gamma$ in the minimally extended model described previously. The one loop diagrams inducing these decays
are shown in Figures \ref{fig:tri} and \ref{fig:triS}.   There are two classes of diagrams, one shown in Figure \ref{fig:tri} in which the fermions in the loop do not change identities, which we refer to as flavor-conserving ones, and the other as shown in Figure \ref{fig:triS} in which the fermions in the loop change identities which we refer to as flavor-changing  diagrams.  $h\to \gamma\gamma$ receives contributions from the flavor-conserving class, and $h\to Z \gamma$  receives   contributions from the  both classes. These features make it possible to have NP contributions of the new fermions to $h \to \gamma \gamma$ and $h \to Z \gamma$ differently to addressing the problem we are dealing with.

Depending on the original parameters $m_D$, $m_S^X$ and $c_f^X$, there are three classes of possible mass eigenstates: 1. both $m_{1,2}^X$ are positive; 2. one of the $m^X_{1,2}$ is positive and the other is  negative; and 3. both $m^X_{1,2}$ are negative. 
In the last case, one can perform a chiral rotation \(e^{i \gamma_5\pi/2}\) on the fields \(f^X_{1,2}\) to make all masses positive and transform \({\cal L}_h \to - {\cal L}_h\), while leaving the other interaction terms unchanged. This would not change the final outcome comparing to the first case, as our final result depends on the value of $(c_f^{X})^2$. Therefore, we only need to consider the first two  possibilities. These two cases can have different features  for  $h\to \gamma\gamma$ and $h \to Z \gamma$. We proceed to discuss them in the following.

\subsection{ The case for $m^X_{1,2}$ both to be positive}

The Feynman diagrams with flavor-conserving vertices are depicted in Figures~\ref{fig:tri} . The calculations are similar to those in the SM. 
The  contributions with $f$ in the fermion loop are  given as 
\begin{equation}\label{conserv}
	\delta 	c_Z^{f} =   Q_f  g_h^f   v
 g_Z^f    
\frac{1}{m_f}	I_Z(\tau_f,\lambda_f)\,,
	~~~	\delta c_\gamma  ^f   =      Q_f^2   g^f_h v \frac{1}{m_f} A_f(\tau_f)  \,, 
\end{equation}
where 
$f\in \{ f_{1}^Y,f_{2}^Y , f_{1}^{Y+1}, f_{2}^{Y+1}\}$ and 
 $\tau_f  = m_h^2/4m _f^2, ~\lambda_f= m_Z^2 / 4 m_f^2 $. Here $Q_f$, $g_h^f$ and $g_Z^f$ are the coupling strengths of $A^\mu$, $h$ and $Z^\mu$ to $f$, where $Q_{f_{1,2}^X}=X$, $g_h^{f^X_1}=\sin 2 \theta_X c_f^X   /\sqrt{2}$,  
 $g_h^{f^X_2}=-  \sin 2 \theta_X  c_f^X   /\sqrt{2}$,
 $g_Z^{f^X_1}=- (\eta_X(1 + \cos 2 \theta_X) +4 X \sin^2  \theta_W)/4\cos \theta_W \sin \theta_W$, and  $
 g_Z^{f^X_2}=- (\eta_X(1 - \cos 2 \theta_X) +4 X \sin^2  \theta_W)/4\cos \theta_W \sin \theta_W$.
 The loop functions are given as ~\cite{Ellis:1975ap,Cahn:1978nz,Shifman:1979eb,Gavela:1981ri,Bergstrom:1985hp}
\begin{equation}
	\label{eq:SMloopfuncts}
	\begin{split}
		I_Z(a,b) &  =   \frac{2 }{(a-b)}+\frac{2+2b-2a}{ (b-a)^2} \, (f(b)-f(a)) +\frac{4b}{(b-a)^2} \, (g(b)-g(a))   \,, \\[0.15cm]
		A_f(\tau)&=  \frac{2}{\tau^2} \left [ (\tau-1)f(\tau)  + \tau \right ]\,,\\
		f(\tau)		&=   \left\{ \begin{array}{lll}
			{\rm arcsin}^2\sqrt{\tau} & \tau \le 1 \\ -\frac{1}{4}\left(\log\frac{\sqrt{\tau}+\sqrt{\tau-1}}{\sqrt{\tau}-\sqrt{\tau-1}}-i\pi \right)^2 & \tau > 1 \end{array}\right. \,,\\
		g(\tau) & =
		\begin{cases}
			\sqrt{1/\tau-1} \,\arcsin (\sqrt{\tau}) &  \tau \leq 1 \\[0.1cm]
			\frac{1}{2}\sqrt{1-1/\tau}\l(\log\frac{\sqrt{\tau}+\sqrt{\tau-1}}{\sqrt{\tau}-\sqrt{\tau-1}}-i \pi\r)   & \tau>1 
		\end{cases}\,. \\
	\end{split}
\end{equation}
At $m_f \gg m_{h,Z}$, we have  
$I_Z(\tau_f , \lambda_f) = A_f (\tau_f) = 4/3$. 

The Feynman diagrams with flavor-changing vertices are depicted in Figure~\ref{fig:triS}. 
Due to the Ward identity, the photon vertices must conserve the flavor and hence 
this type of diagram is absent in $h\to \gamma \gamma$. 
The contribution of the $f^X$ doublet to $\delta c_Z$ is 
\begin{eqnarray}\label{10}\label{flavor-vio}
	\delta 	c_Z ^ {f_1^Xf^X_2} &=& 
	v  Q_f     g_h ^{f^X_1 f^X_2}  g_Z ^{f_1 ^X f_2^X }
	L(m^X_1,m^X_2)  \,,
\end{eqnarray}
with the loop function  given as 
	\begin{eqnarray}
		L(m_1,m_2) 
		&=&
		8\frac{m_1}{m_h^2 - m_Z^2} + 
		8 \frac{m_1}{(m_h^2 -m_Z^2)^2}
		m_Z^2 \left(
		B_0(m^2_h ,m_1,m_2) -
		B_0(m^2_Z ,m_1,m_2) 
		\right) \nonumber\\ &+& 
		4m_1 \left[ 
		\frac{ 2 m_1 ( m_1+m_2)}{m_h^2 - m_Z^2} -1 
		\right]
		C_0 (0, m_h ^2, m^2_Z , m_1,m_1, m_2) + (m_1 \leftrightarrow m_2) \,.
	\end{eqnarray}
Here
	$B_0$ and $C_0$ are the Passarino-Veltman functions~\cite{Passarino:1978jh, Patel:2015tea} and their analytical forms can be obtained by Package-X~\cite{Patel:2015tea}.
The coupling strengths can be read off from Eq.~\eqref{Lag} as 
$g_h^{f_1^Xf^X_2} = c_f^X   \cos 2 \theta_X /\sqrt{2}$ 
and $g_Z^{f_1^X  f_2^X} = \eta_X  \sin 2 \theta_X / ( 4 \cos \theta_W \sin \theta_W)$. 

After carrying out the four-momentum integral, we arrive at 
\begin{eqnarray}\label{13}
	L (m_1,m_2) &=&  2 \int ^1_0 dy  \int ^1_0 y dx 
	\frac{ -m_1 (2 y+1) xy +2 m_1 x^2y^2 +m_1+m_2 xy  (-2 y+2 xy +1)  }{  (1 -xy  ) m_1 ^2 
		+	 xy  m_2 ^2  
		- xy  m_h^2   + x^2y^2  m_h^2 + (m_h^2 - m_Z^2)xy (1-y) }\nonumber\\
	&+&2 \int ^1_0 dy  \int ^1_0 y dx 
	\frac{ m_1 (2 y-3) xy+m_1+m_2 (2 y-1) xy  }{  		 (1-xy)m_1^2 
		+xy  m_2^2 -(y-xy)(1 - y +xy )  m_Z^2-xy (1-y)m_h^2  } \nonumber \\
	&+& (m_1 \leftrightarrow m_2)\,.
\end{eqnarray}
Since the new fermions are expected to be much heavier than the SM particles,  we  drop  $m_Z^2$ and $m_h^2$ and  obtain 
\begin{eqnarray}\label{heavyf12}
	L (m_1 ,m_2) =
	\frac{2  \left(m_1^2-4 m_1m_2+m_2^2\right)}{(m_1-m_2)^2 (m_1+m_2)}
	-
	\frac{4 m_1m_2\left(m_1^2-m_1m_2+m_2^2\right) 
		\log 
		(m_2^2/m_1^2)
	}{(m_1-m_2)^3 (m_1+m_2)^2}\,.
\end{eqnarray} 
Around  $m_1=  m_2$, we have 
\begin{eqnarray}
	L(m_1,m_2) 
	&=& \frac{16}{3(m_1 + m _2)  } 
+ {\cal O}\left(
\frac{(m_1-m_2)^2}{(m_1+m_2)^3 }
\right) 
  \,.
\end{eqnarray}
The above leads to 
\begin{eqnarray}\label{91}
	\delta c _\gamma 
	&=&-\frac{4}{3}\left( 
	Y^2
	\frac{\left( c_f^Y v\right)^2}{m_1^{Y} m_2 ^Y} 
	+ (Y+1)^2
	\frac{\left( c_f^{Y+1} v\right)^2}{ m_1 ^{Y+1}  m _2 ^{Y+1} } \right)\,, \\
\delta c_Z 
&=&
\frac{4 \sin \theta_W }{3 \cos \theta_W }
\left( 
Y^2
\frac{\left( c_f^Y v\right)^2}{m_1^{Y} m_2 ^Y} 
+ (Y+1)^2
\frac{\left( c_f^{Y+1} v\right)^2}{ m_1 ^{Y+1}  m _2 ^{Y+1} } \right)+
\frac{2}{3 \sin \theta_W \cos \theta_W}\left( 
Y 
\frac{\left( c_f^Y v\right)^2}{m_1^{Y} m_2 ^Y} 
- (Y+1) \frac{\left( c_f^{Y+1} v\right)^2}{ m_1 ^{Y+1}  m _2 ^{Y+1} }
\right), \nonumber
\end{eqnarray}
around $|m_1^X| \approx |m_2^X|$. 
Therefore, for the case with  $m_{1,2}^X>0 $, 
we see that $\delta c_{\gamma}$~($\delta c_Z$) constructively~(destructively) interfere with 
$c_\gamma^{\text{SM}}$~($c_Z^{\text{SM}}$), which would lead to $\mu _\gamma >1$ and $\mu_Z<1$ for $|\delta c_{Z}| < |c_{Z}^{\text{SM}}|$. This trend is maintained for   keeping   $m_h$ and $m_Z$. Hence, the scenario of $m_{1,2}^X>0$ is not able to explain the excess of $\mu_Z^{\text{exp}}$.

\subsection{ The case for one of $m^X_{1,2}$ to be negative}

From Eq.~\eqref{91}  it is observed that by  setting $m_2^{Y+1}<0$, a  destructive interference between the $f^Y$ and $f^{Y+1}$ doublets for $\delta c_\gamma$ can happen while leaving the second term in \(\delta c_Z\) to constructively interfere. To rigorously address the scenario with $m_2^{Y+1}<0$, a rotation of $f_2^{Y+1} \to \gamma_5 f_2^{Y+1}$ is necessary to ensure the positiveness of the fermion mass.
	While ${\cal L}^Y_{M,h,\gamma,Z}$ remain unchanged, the others are modified to 
	\begin{eqnarray}\label{rota}
		{\cal L}^{Y+1}_M
		&=& 
		- m_1^{Y+1}  
		\overline{f}_1^{Y+1} f_1^{Y+1}
		-
		|m_2 ^{Y+1}| 
		\overline{f}_2^{Y+1}  f_2^{Y+1}\,,\nonumber\\
		{\cal L}_h ^{Y+1}
		&=&			- 
		\frac{ c_f^{Y+1} h}{\sqrt{2}}
		\overline{f}^{Y+1}  \left( 
		\sin 2 \theta_{Y+1}  + \cos 2 \theta_{Y+1}  i \sigma_y \gamma_5   
		\right)f^{Y+1}
		\,,\\
		{\cal L}_{Z} ^{Y+1}
		& = &e
		\overline{f} ^{Y+1}
		Z ^\mu \gamma_\mu 
		\left[ 
		- (Y+1)  \tan \theta_W    +  \frac{1 }{4 \cos \theta_W \sin \theta_W }
		\left( 1
		+   \cos 2 \theta_{Y+1}\sigma_z   
		-  
		\sin  2  \theta_{Y+1} \sigma_x  \gamma_5
		\right) 
		\right]
		f^{Y+1} \nonumber \,, \nonumber\\
		{\cal L} _W 
		&=&
		\frac{g}{\sqrt{2}}
		W^+_\mu 
		\Big(
		\cos \theta_{Y+1}  \cos \theta_Y  \overline{f}_1 ^{Y+1} \gamma^\mu  f_1 ^Y
		+
		\sin \theta_{Y+1}  \sin  \theta_Y  \overline{f}_2 ^{Y+1} \gamma^\mu  \gamma_5  f_2 ^Y
		\nonumber\\
		&-& \cos \theta_{Y+1} \sin \theta_Y \overline{f}_1 ^{Y+1} \gamma^\mu  f_2 ^Y
		-
		\cos \theta_{Y} \sin \theta_{Y+1}  \overline{f}_2 ^{Y+1}  \gamma^ \mu  \gamma_5 f_1 ^Y
		\Big) +(\text{h.c.}) \,.\nonumber
	\end{eqnarray}
The Lagrangian is significantly modified with an extra \(\gamma_5\) appearing in the off-diagonal interactions. This shows that the physical quantities depend not only on the absolute values of the masses but also on their signs. This can be traced back to the fact that the mass terms in our model originate from both the bare Lagrangian and the Higgs mechanism, and the two cannot be diagonalized simultaneously with chiral rotations.

We have carried out detailed calculations and find that
for $h\to \gamma \gamma$ and $h \to Z \gamma$, the consequences of this rotation can be effectively modeled by substituting $m_2^{Y+1}$ with $-m_2^{Y+1}$ in Eqs.~\eqref{10}, \eqref{13} and \eqref{91}. It results in the favorable solution with $| \delta c_\gamma|< | c_\gamma^{\text{SM}}|.$ 

Before ending this section, we note that from Eq.~\eqref{91} there is a second set of solutions with $m_{2}^{Y+1}<0$. By taking $c_f^Y=0$, 
$\delta c_\gamma $ and $c _\gamma^{\text{SM}}$ are opposite in sign and 
it is possible to explain the data with the feature of   $\delta c_\gamma \approx -2c_\gamma^{\text{SM}}$.  
In this scenario, \(f_S^Y\) decouples from the other fermions, and it is not necessary to include it in the model. 

It is worth mentioning that Ref.~\cite{Barducci:2023zml} considered the fermions with the same representations as those in Table~\ref{table:1}. In particular, by considering \(m_D, m_S^{Y,Y+1} \to 0\), we can reproduce the results in Ref.~\cite{Barducci:2023zml}. In this case, the only solution is where \(\delta c_\gamma \approx -2c_\gamma^{\text{SM}}\) with \(|Q_f|\) found to be \(1e\) and \(2e\). However, it is more natural to have a solution where \(c^{\text{SM}}_\gamma \gg \delta c_\gamma\).

\section{Numerical results}

\begin{figure}[t]
	\begin{center}
		\subfigure[]{	\includegraphics[width=0.3 \linewidth]{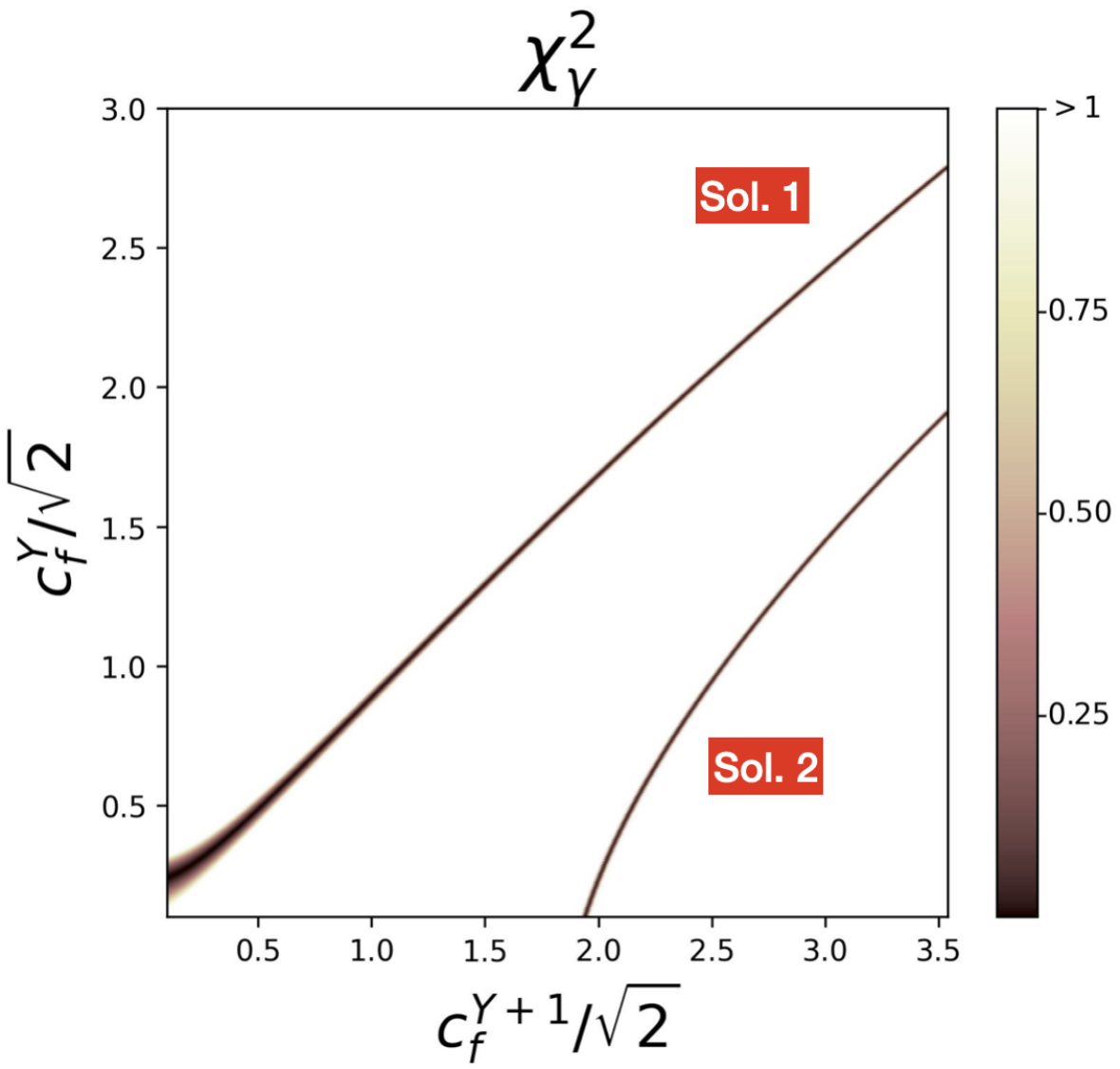}}
		\subfigure[]{\includegraphics[width=0.3 \linewidth]{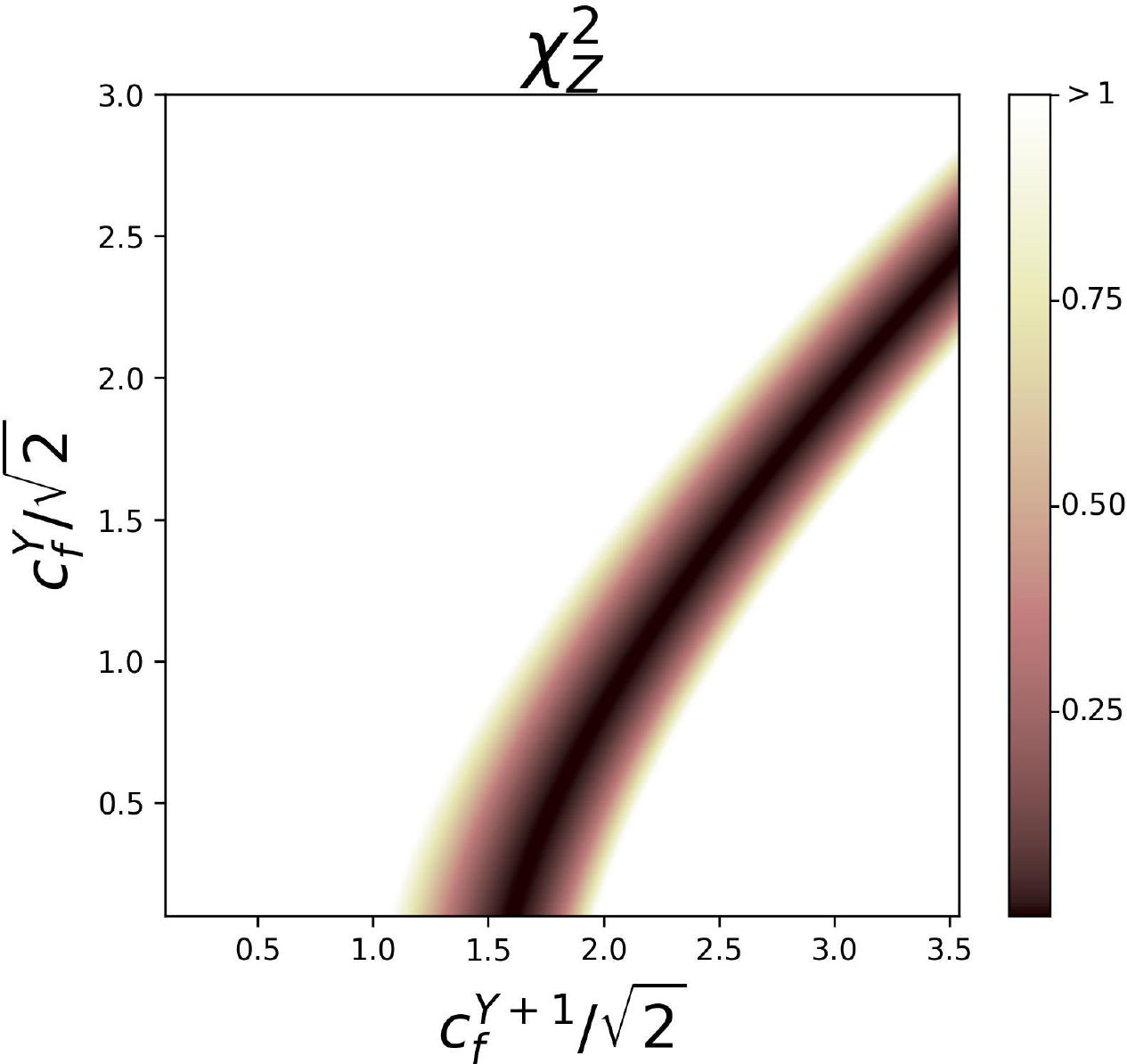}}
		\subfigure[]{\includegraphics[width=0.27 \linewidth]{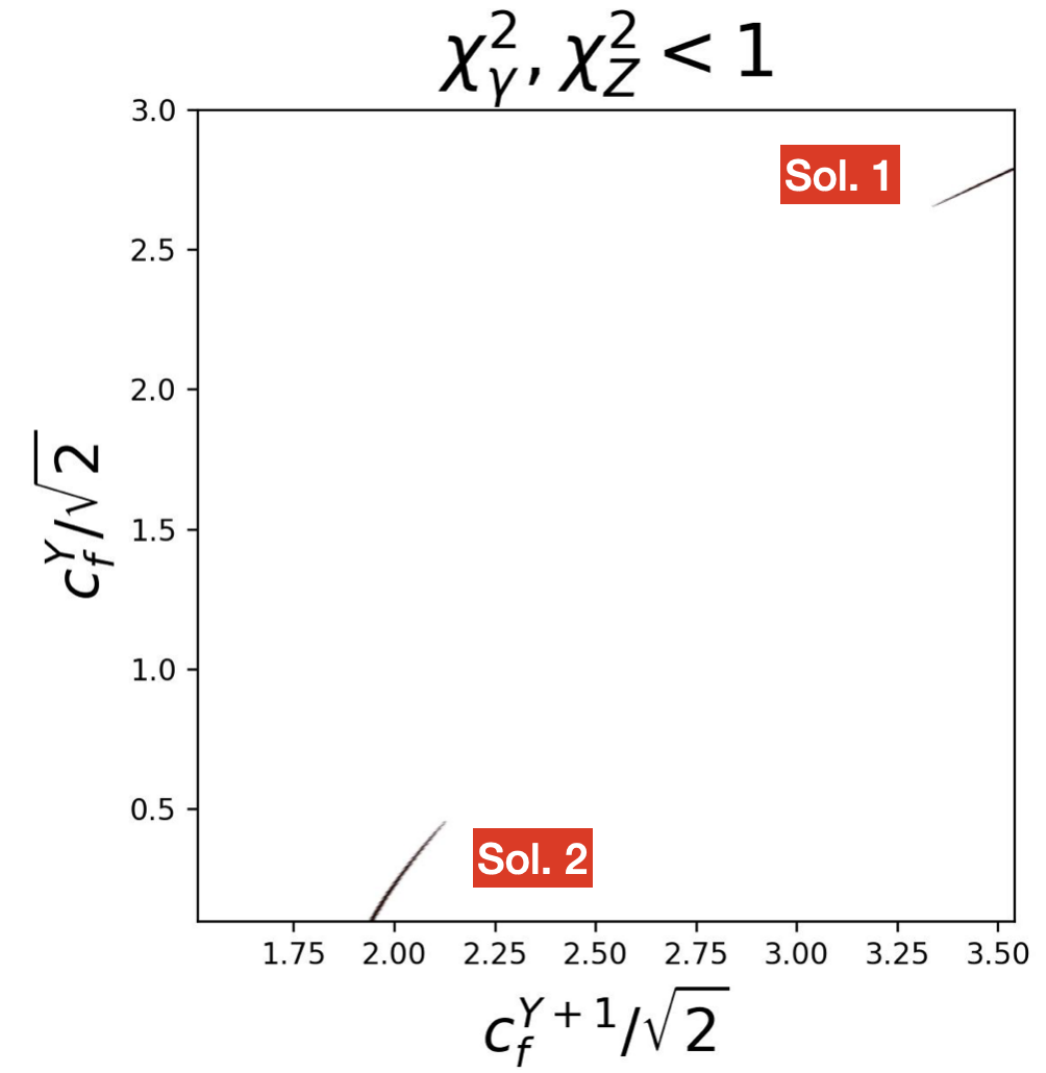}}
		\caption{ In Figures (a) and (b), the colored regions represent the parameter space consistent with \(\mu_\gamma^{\text{exp}}\) and \(\mu_Z^{\text{exp}}\) within one standard deviation, where the color indicates the values of \(\chi^2_{\gamma,Z}\) as defined in Eq.~\eqref{chi2}. Figure (c) displays the regions that satisfy both \(\chi_\gamma^2<1\) and \(\chi_Z^2<1\). 
			In Figures (a) and (c),
			the upper and lower lines  denote the Solutions 1 and   2,  where \(|\delta c_\gamma| \ll |c_\gamma^{\text{SM}}|\) and \(\delta c_\gamma\approx -2c_\gamma^{\text{SM}}\), respectively.}
		\label{fig:two}
	\end{center}
\end{figure}

We now provide numerical results for the case with negative $m_2^{Y+1}$. 
For simplicity, we adopt   
$m_D = m_S^Y = -m_S ^{Y+1} \,,$ and  the masses are given by 
\begin{eqnarray}
	m_{1,2} ^Y &=& m_D  \pm  \frac{c_f^Y v}{\sqrt{2}} \,,~~~
	m_{1} ^{Y+1}   = - m_{2} ^{Y+1}  = \sqrt{ m_D^2 + \frac{ ( c_f^{Y+1}  v)^2}{2}} \,. 
\end{eqnarray}
Without loss of generality, we take $m_1^Y>m_2^Y$. 
Hence, we have the hierarchy of $m_1^Y> | m_{1,2}^{Y+1}| > m_2^{Y}$, making $f_2^Y$ being the stable particle due to the energy conservation.
For a meaningful perturbative calculation, we consider only the regions where $(c_f^X/ \sqrt{2})^2 < 4\pi$  and fix  $Y=-9$.
We also confine the model to have the lightest new charged fermion mass be larger than \(1600\)~GeV to satisfy the experimental lower bound for \(|Q_f|\)  up to 7~\cite{ATLAS:2023zxo}.

To find the regions of parameter which fit data well, we define 
\begin{equation}\label{chi2}
	\chi_\gamma^2 = 
	\left(  \frac{\mu^{\text{exp}} _\gamma - \mu_\gamma  }{\sigma^{\text{exp}} _\gamma}
	\right)^2 
	\,,~~
	\chi_Z^2 =    \left(  \frac{\mu^{\text{exp}} _Z - \mu_Z  }{\sigma^{\text{exp}} _ Z}\right)^2 \,,
\end{equation}
where $\sigma^{\text{exp}}_{\gamma,Z} $ stand for the experimental uncertainties of  $\mu_{\gamma,Z }$. 
In Figure~\ref{fig:two}(a), (b) and (c), we plot the allowed parameter spaces by setting  $\chi_\gamma^2<1$, $\chi^2_Z<1$ and $\chi_\gamma^2,  \chi_Z^2<1$, respectively. 
From Figure~\ref{fig:two}(a) and (c), it is observed that there are two sets of solutions. 
We name the upper line Solution 1, characterized by $|\delta c_\gamma| \ll |c_\gamma^{\text{SM}}|$, while the lower line is named Solution 2, characterized by $\delta c_\gamma \approx -2c_\gamma^{\text{SM}}$, representing the solution mentioned at the end of the last section.
For $\chi^2_Z$ depicted in 
Figure 3(b), there is  only one set of  solutions with $|\delta c_Z |<|c_Z^{\text{SM}}|$  in contrast.  It is worth noting that, if we consider $m_{2}^{Y+1} > 0$, there would be no solution here. In the limit of $m_D \gg c_f^X v$, the lines in Figure~\ref{fig:two} exhibit a slope of $ (Y+1)^2/Y^2 $, necessary to achieve a cancellation between the $f^Y$ and $f^{Y+1}$ doublets in Eq.~\eqref{91}.

We also show in Figure~\ref{fig:two} (c), the two sets of solutions with the smallest couplings. They  are 
\begin{eqnarray}\label{parameters1}
&&\mbox{Solution 1}:\;\;\;\; \left( c_f^{Y+1} ,  c_f^{Y} \right)_{\delta c _\gamma \approx 0}  = \left(  4.73,~3.80 \right) \,,~~~\nonumber\\
&&\mbox{Solution 2}: \;\;\;\;\left( c_f^{Y+1} ,  c_f^{Y} \right)_{\delta c _\gamma \approx -2 c_\gamma^{\text{SM}} }= ( 2.69, 0 ) 
\,.
\end{eqnarray}
The  corresponding predictions  for other parameters are given by
\begin{eqnarray}\label{parameters2}
\mbox{Solution 1}&:&\;\; \left(
\theta_{Y+1} , \theta_Y 
\right) = (10^\circ , 45 ^\circ ),\;\;\;\;\left(
\mu_\gamma , \mu_Z 
\right)  =
(1.10, 1.55) \nonumber\\
&&\;\;
( m_1^Y, m_2^Y)
=(2923, 1600)~\text{GeV}\,,~~~~
 m_{1,2}^{Y+1} 
=2407  ~\text{GeV}\,,
\nonumber\\
\mbox{Solution 2}&:&\;\;(\theta_{Y+1},\theta_Y ) 
=  (8^\circ,   0 ^\circ ),\;\;\;\;
\left(
\mu_\gamma , \mu_Z 
\right) =
(1.10, 2.88) \,,\nonumber\\
&&\;\;
( m_1^Y, m_2^Y)
=(1600, 1600)~\text{GeV}\,,~~~~
m_{1,2}^{Y+1} 
=1667~\text{GeV}\,.
\end{eqnarray}
We therefore have found two classes of solutions   modify $\mu_\gamma$ slightly, but enhance  $\mu_Z$  to the current average value. Solution 2 seems to be a contrived solution although cannot be ruled out simply using branching ratio measurements. 
If the current data are confirmed, we would think Solution 1 to be a better solution.
It is interesting to point out that due to the large  hypercharges introduced, we would encounter the Landau pole in the \( U(1)_Y \) gauge coupling around 10 TeV in both solutions.  If the excess is  confirmed,
our model will imply  additional NP around 10 TeV, which may be detected by future 
  experiments with higher precision and energy.
 It will be interesting to study which kinds of additional NP we need to remedy the situation.

Due to the smallness of the radiative corrections, we expect the large cancellation in Solution 1 to also occur after including the higher-order corrections. For instance, we consider the next-to-leading order (NLO) QED correction.
In the $m_{1,2}^{Y,Y+1} \gg m_h$ limit, it shifts the amplitudes by
\begin{equation}
\delta c_{Z,\gamma}^{f} \to 
\left( 1 - \frac{3}{4} Q_f^2  \frac{\alpha_{em}}{\pi}  
\right) \delta c_{Z,\gamma}^{f}, ~
\delta c_Z^{f_1^X f^X_2} \to 
\left( 1 - \frac{3}{4} Q_f^2  \frac{\alpha_{em}}{\pi} 
\right) \delta c_Z^{f_1^X f^X_2}.
\end{equation}
The formalism can be cross-checked by replacing  $Q_f^2 \alpha_{em}$ with $C_F \alpha_s$  and comparing to the NLO QCD corrections~\cite{Zheng:1990qa,Spira:1991tj}, as both have the same topological diagrams at NLO. Since the amplitudes are shifted homogeneously, we conclude that the large cancellations also occur at NLO.
 

Before ending the discussion, we would like to comment on some other phenomenological aspects  at colliders. 
The new fermions in the model can be produced which has been discussed for smaller $|Y|$ and masses in Ref.~\cite{Barducci:2023zml}. Since in our model the hypercharge $|Y|$ is  large, the lightest new fermion is constrained to have a mass larger than $1600$~GeV~\cite{ATLAS:2023zxo}. The production at the current LHC may be scarce or absent. 
However, with higher energies and higher luminosity, the new fermions in our model may be produced. The signature of the lightest fermion will leave a charge track in the detector to be measured. While to the heavier ones, if produced they can decay into other final states.
For Solution 1, except for $f_2^Y$ the others will  
decay into $f_2^Y$  plus either a $W$-boson or $Z$-boson.  But these decays will have large widths of the order of  hundreds GeV, making detection difficult since there will not be a sharp resonance peak to look for. On the other hand, for Solution 2, the fermion mass differences are not high enough to form an on-shell gauge boson. They decay into an off-shell 
$W$-boson, which then decays into a charged lepton and a neutrino, with the decay widths being on the order of  a few tens of MeV.  Similarly, the decay widths to light quark jet pairs are twice as large.  Such measurements may provide information to distinguish between the two   solutions.

The signal strengths for \(h \to ZZ^*\) (\(\mu_{ZZ}\)) and \(h \to WW^*\) (\(\mu_{WW}\)), normalized to the  SM  predictions, are measured to be \(1.02 \pm 0.08\) and \(1.00 \pm 0.08\)~\cite{ParticleDataGroup:2022pth}, respectively. In the SM, these processes are predominantly driven by tree-level couplings, while our model influences them at the NLO  through triangular fermionic loops. For both Solutions 1 and 2, the  effects on \(\mu_{WW}\) are highly suppressed by $(v/m_{1,2}^X)$ and $\alpha_{em}$, with corrections below the one percent level. 
The correction to \(\mu_{ZZ}\)
suffers the same suppression but 
 is amplified by the large $Y$. However, in Solution 1, the  destructive mechanism responsible for the small \(\delta c_\gamma\) occurs also in  \(\mu_{ZZ}\), leading to   correction    below one percent. In contrast, Solution 2 delivers the largest correction, with \( \mu_{ZZ}  -1 = 3 \% \), but is still much smaller than the current experimental precision of $8\%$.
To probe the footprints of NP and CP violation, it would be useful to study the \( p_T \) spectrum of cascade decays of gauge bosons. A comprehensive analysis  should be carried out once future experiments achieve the required precision.

\section{Conclusion}

We have studied a possible explanation for the \(h \to Z \gamma\) excess observed in recent measurements by ATLAS and CMS through the construction of a renormalizable model.
If this excess is confirmed, it would be a signal of  NP beyond the SM. 
For NP contribution, while modifying $h\to Z \gamma$, in general it  also   affects $h \to \gamma \gamma$. Since the measured $\mu_\gamma$ agrees well with the SM prediction, there are tight  constraints on theoretical models trying to simultaneously explain data on $h\to \gamma\gamma$ and $h\to Z \gamma$. 
We find that a minimally fermion singlets and doublet extended NP model can explain simultaneously the current data on these decays. We have identified two solutions. One is the SM amplitude $c_Z^{\text{SM}}$ is enhanced by $\delta c_Z$ for $h \to Z \gamma$ to the observed value, but the $h \to \gamma \gamma$ amplitude $c_\gamma^{\text{SM}} +\delta c_\gamma$ is modified  to $-c_\gamma^{\text{SM}}$ to give the observed branching ratio. This seems to be a contrived solution, although it cannot be ruled out simply using branching ratio measurements. We, however, have found  another solution which naturally enhances the $h \to Z \gamma$ to the measured value, but keeps the $h \to \gamma\gamma$ close to its SM prediction.  With high energy colliders which can produce these new heavy fermions, by studying the decay patterns of the heavy fermions, it is possible to distinguish the two solutions we found. We eagerly await future data to provide more information.

\begin{acknowledgments}

This work is supported in part by 
the National Key Research and Development Program of China under Grant No. 2020YFC2201501, by
the Fundamental Research Funds for the Central Universities, by National Natural Science Foundation of P.R. China (No.12090064, 12205063, 12375088 and W2441004). 
\end{acknowledgments}

\appendix\section{ General Yukawa interaction structure in the model}

In this appendix, we provide some details for a  more general form for 
Eq.~(\ref{Model}) in our model. The Yukawa interaction is given by
\begin{equation}
	\small
\begin{aligned}
	\mathcal{L}_{H+M}=&-m_D\overline{f}_D f_D
	-m_S ^{Y} \overline{f}_S ^Y  f_S ^Y 
	-m_S ^{Y+1}\overline{f}_S ^{Y+1}  f_S ^{Y+1}  
	- \left(c_f^Y\overline{f}_D e^{i\alpha^Y \gamma^5} f_S^{Y}H
	+c_f^{Y+1}\overline{f}_D e^{i\alpha^{Y+1} \gamma^5}f_S^{Y+1} \tilde{H}
	+(\text{h.c.})\right)\;.
\end{aligned}
\end{equation}
We note that although it is possible to rotate away   $\alpha^X$ by the chiral rotations of $f_S^{X} \to e^{-i\alpha^X\gamma_5} f_S^X$, the mass terms would acquire chiral phases of $e^{-2i\alpha^X \gamma_5}$, which violates CP symmetry. Hence, $\pi > |\alpha^X| > 0$ would necessarily lead to a CP-violating theory. 
One can apply a chiral rotation to the Lagrangian to shift phases between different terms without affecting the physical results. We choose, without loss of generality, a chiral basis in which the chiral phases vanish in the mass terms.

After spontaneous symmetry breaking, 
we could diagonalize these mass matrices by biunitary transformation,
\begin{equation}
	\begin{aligned}
		\begin{pmatrix}
			f_{D,L}^{X}\\ f_{S,L}^{X}
		\end{pmatrix}
		\Rightarrow 
		V_{L}^{X}
		\begin{pmatrix}
			f_{1,L}^{X}\\ f_{2,L}^{X}
		\end{pmatrix},\;
		\begin{pmatrix}
			f_{D,R}^{X}\\ f_{S,R}^{X}
		\end{pmatrix}
		\Rightarrow V_{R}^{X}
		\begin{pmatrix}
			f_{1,R}^{X}\\ f_{2,R}^{X}
		\end{pmatrix},\;
	\end{aligned}
\end{equation}
which leads to 
\begin{equation}\label{A3}
	 (V_{L}^{X})^\dagger 
	 \hat M^X
	 V_{R}^{X}  = 
	 (V_{L}^{X})^\dagger\begin{pmatrix}
	\begin{smallmatrix}
		m_D&\frac{c_f^{X} v}{\sqrt{2}}e^{i\alpha^X}\\
		\frac{c_f^X v}{\sqrt{2}}e^{i\alpha^X}&m_S^{X}
	\end{smallmatrix}
\end{pmatrix}V_{R}^{X}=\begin{pmatrix}
	m_1^{X}&0\\
	0&m_2^{X}
\end{pmatrix}\;.
\end{equation}
	
 We   parameterize $V_{L}^{X}$ as
\begin{equation}
		V_L^{X}=
		\begin{pmatrix}
			1 & 0 \\
			0 & e^{i \psi_{L}^{X}}
		\end{pmatrix}
		\begin{pmatrix}
			\cos \theta_{L}^{X} & -\sin \theta_{L}^{X} \\
			\sin \theta_{L}^{X} & \cos \theta_{L}^{X}
		\end{pmatrix}
		\begin{pmatrix}
			e^{i \eta_{L1}^{X}}&0\\
			0&e^{i \eta_{L2}^{X}}
		\end{pmatrix}\,,
\end{equation}
while 
 $V_{R}^{X}$
 is parameterized  similarly
  by $\psi_{R}^{X}$, $\theta_{R}^{X}$, $\eta_{R1}^{X}$ and $\eta_{R2}^{X}$. 
  Solving Eq.~\eqref{A3}, we arrive at 
\begin{equation}
	\begin{aligned}
			\tan\psi_L^{X}=-
			\tan \psi _R^X = 
			\frac{m_D-m_S^{X}}{m_D+m_S^{X}}\tan\alpha^{X}\,,~~~
			\tan2\theta_{L,R} ^{X}=\frac{\sqrt{2}c_f^{X} v
				\Big(m_D\cos(\psi_L^{X}-\alpha^{X})+m_S^{X}\cos(\psi_L^{X}+\alpha^{X})\Big)
			}{m_D^2-(m_S^{X})^2}\,.
	\end{aligned}
\end{equation}
 Without loss of generality,  we can set $\eta_{R1}^{X}=\eta_{R2}^{X}=0$ and derive that
\begin{equation}
	\begin{aligned}
		&\eta_{L1}^{X}=\arg(m_D\cos^2\theta_{X}+m_S^X e^{-2i\psi_L^X}\sin^2\theta_{X}
		+\frac{c_f^{X} v}{\sqrt{2}}e^{i(\alpha^X-\psi_L^X)}\sin2\theta_{X})\\
		&\eta_{L2}^{X}=\arg(m_S^X e^{-2i\psi_L^X}\cos^2\theta_{X}+m_D\sin^2\theta_{X}
		-\frac{c_f^{X} v}{\sqrt{2}}e^{i(\alpha^X-\psi_L^X)}\sin2\theta_{X})\;.
	\end{aligned}
\end{equation}
We stress  there are only two independent CP phases  $\alpha^Y$ and  $\alpha^{Y+1}$, and $\psi_L^{X}$, $\eta_{L1}^{X}$ and $\eta_{L2}^{X}$ are functions of  them. 

The Lagrangian responsible for the couplings are then given as 
\begin{equation}
	\label{general int}
\begin{aligned}
	\mathcal{L}_{h}^X=&-\frac{c_f^{X} h}{\sqrt{2}}\overline{f}^X 
	\begin{pmatrix}
		e^{i\delta_{1}^{X}\gamma^5}&0\\
		0&-e^{i\delta_{2}^{X}\gamma^5}
	\end{pmatrix}\sin2\theta_X f^{X}
	+\left(\frac{c_f^{X} h}{\sqrt{2}}\cos2\theta_X
	\overline{f}_{1}^X e^{i\frac{\delta_{1}^{X}+\delta_{2}^{X}}{2}\gamma^5}
	f_{2}^X
	+\text{h.c.}\right)\\
	\mathcal{L}_{Z}^{X}=&-eZ ^\mu\overline{f} ^X  \gamma_\mu 
	\left[X \tan \theta_W+\tfrac{\eta_X}{2 \sin 2\theta_W }
	\left(1+\cos 2 \theta_X  \sigma_z\right) 
	\right]f^X
	+eZ ^\mu\eta_X\tfrac{\sin 2 \theta_X}{2 \sin 2\theta_W }
	\Bigg(\overline{f}_{1} ^X  \gamma_\mu 
	e^{i\frac{\delta_{2}^{X}-\delta_{1}^{X}}{2}\gamma^5}f_2^X+\text{h.c.}\Bigg)
	\;,
\end{aligned}
\end{equation}
where $\delta_{1}^{X}=\alpha^X-\psi_{L}^{X}-\eta_{L1}^{X}$ and $\delta_{2}^{X}=\alpha^X-\psi_{L}^{X}-\eta_{L2}^{X}$. 
In the limit of  $\delta_{1}^{X}=\delta_{2}^{X}=0$, they reduce to  Eq.(\ref{Lag}).

If $\alpha^X$ is non-zero, there will be two types of interactions contribute to $h \to Z \gamma$, one is CP-conserving proportional to $h Z_{\mu \nu}F^{\mu \nu} $ and another CP-violating to $h Z_{\mu \nu } \tilde F^{\mu \nu}h$. 
For illustration, let us fine-tune $\alpha^X$ so that   $\delta_1^X = \delta_2^X = \delta_{CP}$. 
Taking $m_{1,2}^X \gg  m_h$,
the 
effects of the novel fermions can be encapsulated by the 
effective Lagrangian of 
\begin{equation}\label{newminwey}
	\mathcal{L}_{\mathrm{eff}}^{hZ \gamma}
	=\frac{\alpha_{em}}{4 \pi  v}(c_{Z}^{\text{SM}}+\delta c_{Z}\cos\delta_{\mathrm{CP}})
	Z_{\mu \nu}F^{\mu \nu}h
	+\frac{\alpha_{em}}{4 \pi  v}\delta\tilde{c}_{Z}\sin\delta_{\mathrm{CP}}
	Z_{\mu \nu}\tilde{F}^{\mu \nu}h\;,
\end{equation}
with  $\tilde{F}^{\mu \nu}=\varepsilon^{\mu\nu\alpha\beta}F_{\alpha\beta} /2 $. At the limit of $m_{1,2}^X \gg ( |m_1 ^X| - | m_2^X | )/2$,  
we have  a simple relation 
$\delta \tilde c _Z = -3  \delta c_Z /2 $  with  
$\delta c_Z$  given by  Eq.~\eqref{91}. 
The  signal strength  of $h\rightarrow Z\gamma$ is modified to  be
\begin{equation}
\mu_Z = 1 + \left[ 
2 \frac{\delta c_Z }{   c_Z ^{\text{SM} }}   \cos \delta _{\text{CP}} 
+
\left(
 \frac{\delta c_Z }{   c_Z ^{\text{SM} }}  
\right)^2  \left(  1+    \frac{5}{4} \sin^2 \delta _{\text{CP}}
\right)
\right]\,.
\end{equation}
The current experiments do not reach the precision required for testing CP violation in \(H \rightarrow Z\gamma\). For simplicity, we took  \(\delta_{CP} = \alpha^X = 0\) in Eq.~\eqref{Model}.

\end{document}